# Experimental Realization of Multiple Topological Edge States in a One-Dimensional Photonic Lattice

*Zhifeng Zhang[1], Mohammad Teimourpour[2], Jake Arkinstall[3], Mingsen Pan[4,5], Pei Miao[4,5], Henning Schomerus[3], Ramy El-Ganainy[2†], and Liang Feng[5*]*

*,†Corresponding Author: E-mail: *fenglia@seas.upenn.edu (L.F.); †ganainy@mtu.edu (R.E.)

[1]Department of Electrical and Systems Engineering, University of Pennsylvania, Philadelphia, PA 19104, USA
[2]Department of Physics and Henes Center for Quantum Phenomena, Michigan Technological University, Houghton, MI, 49931, USA
[3]Department of Physics, Lancaster University, Lancaster, LA1 4YB, United Kingdom
[4]Department of Electrical Engineering, The State University of New York at Buffalo, Buffalo, NY 14260, USA
[5]Department of Materials Science and Engineering, University of Pennsylvania, Philadelphia, PA 19104, USA

**Abstract:** Topological photonic systems offer light transport that is robust against defects and disorder, promising a new generation of chip-scale photonic devices and facilitating energy-efficient on-chip information routing and processing. However, present quasi one-dimensional designs, such as the Su-Schrieffer-Heeger (SSH) and Rice-Mele (RM) models, support only a limited number of nontrivial phases due to restrictions on dispersion band engineering. Here, we experimentally demonstrate a flexible topological photonic lattice on a silicon photonic platform that realizes multiple topologically nontrivial dispersion bands. By suitably setting the couplings between the one-dimensional waveguides, different lattices can exhibit the transition between multiple different topological phases and allow the independent realization of the corresponding edge states. Heterodyne measurements clearly reveal the ultrafast transport dynamics of the edge states in different phases at a femto-second scale, validating the designed topological features. Our study equips topological models with enriched edge dynamics and considerably expands the scope to engineer unique topological features into photonic, acoustic and atomic systems.

The mathematical field of topology, which deals with quantities that preserve their values during continuous deformation, has firmly emerged as a new paradigm for describing new phases of matter since its first applications to condensed matter systems over three decades ago.[1-3] Due to the mathematical equivalence between the paraxial wave equation describing the propagation of light and the Schrödinger equation for the time-evolution of electrons [4], topological concepts seamlessly transfer into the realm of optics and photonics.[5] This realization has inspired a range of versatile topological photonic platforms based on optical resonator arrays,[6-8] waveguide array lattices,[5, 9, 10] photonic crystals,[11-17] and optical quasicrystals.[18, 19] Novel topological features such as symmetry-protected interface states promise a new generation of robust, defect-tolerant and scattering-free photonic circuits[20, 21] with direction-dependent beam dynamics. More recently, a variety of topological lasers have been developed in both one-dimensional (1D) [22-24] and two-dimensional (2D) configurations[25, 26], which provide robust and highly efficient laser actions.

To date, the Su-Schrieffer-Heeger (SSH) Hamiltonian[27] serves as an archetypical model for describing topological physics and designing practical structures, especially in one dimension. However, the topological features of most conventional models are limited to only two dispersion bands, thereby permitting only a limited range of topological numbers characterizing the bands and gaps and consequently restricting the accessible nontrivial phases. Much can be gained from richer models with a larger range of nontrivial phases that can be manipulated systematically to realize the formation of independent topological states. While novel topological phases have been observed with time-periodic driving systems,[5] such systems require unique three-dimensional fabrication techniques that are challenging to be applied for on-chip integrated photonics. Here, we successfully demonstrate the formation of topological edge states associated with multiple bandgaps in a discrete photonic lattice based on standard silicon fabrication techniques. Our system consists of a versatile waveguide array requiring only a small number of fundamental components, and is guided by the concept of generating topological effects through strategic rearrangements that break some crystal symmetries.[28] By varying the design parameters of the waveguides from one photonic lattice to another, we observe a topological phase transition from a regime with a single edge state to a regime with two such states at the same edge. These phases are experimentally distinguished by their different localization and diffraction patterns, and further confirmed by their ultrafast transport dynamics at the femtosecond scale. The coexistence of multiple states at a single edge results in intriguing edge dynamics which allows it to be validated by a characteristic spatial beating effect. Based on the powerful universality of topological concepts, these findings can be directly transferred to a wide range of platforms, such as quantum-optical, acoustic, polaritonic and atomic systems.

The conceptual basis of our investigation starts with a two legged ladder system with two sites per unit cell as shown in the top panel of **Figure 1(a)**. While such a system possesses two dispersion bands with only a single band gap in between, the associated topological features can be further enriched by taking a nontrivial square root[28] of the orignal system to expand the two dispersion bands into four. This transforms the two bands into four bands arranged symmetrically at positive and negative energies, which become associated with a symmetry-reduced tight-binding system with four sites per unit cell. The latter can be represented as a linear bowtie chain with nearest-neighbor couplings, as shown in the bottom panel of Figure 1a. This bowtie structure can be interpreted as a variant of the ubiquitous SSH[27] and Rice-Mele (RM)[29] models. To experimentally probe the topological features of the proposed structure, we investigate a photonic implementation based on an array of coupled waveguides where each waveguide represents a site in the bowtie chain, fabricated on a silicon-on-insulator (SOI) platform as illustrated in **Figure 1(b)**.

In our waveguide photonic lattice system, the intentially designed weak coupling coefficients between neighboring waveguides result in a discrete diffraction length of tens of μm (i.e. the length needed for light to couple completely from one waveguide to its adjacent waveguide), which is much larger than the operation wavelength of approximately 1550 nm. As a result, the paraxial approximation is valid here.[30] Together with the absence of reflection in the propagation direction and orthogonality of all waveguide modes, the Hamiltonian formalism can be safely applied to our system. In anticipation of our experimental results we use notations from the coupled-mode theory, where the propagation constants of the waveguides are denoted by $\beta_n$, precisely controlled by designing the dimensions of the cross section of the waveguides, and the coupling of adjacent waveguides is denoted by coefficients $\kappa_n$, effectively tuned by the distances between the adjacent waveguides. Each waveguide supports only a single fundamental quasi-TM mode. The loss of the Si waveguides is negligible at the operation wavelength compared to the site energy and coupling strength, so that we can safely assume the propagation constants to be real. The lattice structure can then be viewed as a collection of coupled dimers, where $\tilde{\kappa}$ and $\kappa$ indicate the coupling between two sites inside the same dimer, and the coupling between the edge elements of two different neighboring dimers, respectively. The dimers are arranged in an alternating fashion in two orientations that we denote as R and L (Figure 1(a)). The lattice chain is designed to be $(RL)_n R$, with $(RL)$ for the unit cell. The dynamics of the bowtie chain is then described by the evolution equations

$$i\frac{d\vec{a}_n}{dz} = H\vec{a}_n + T\vec{a}_{n-1} + T^\dagger \vec{a}_{n+1}, \qquad (1)$$

where $z$ is the propagation distance along the waveguides and the vectors $\vec{a}_n$ correspond to the field amplitudes of the four waveguides (sites) within the $n$th unit cell. The coupling of two sites in each dimer is described by the intra-cell matrix $H$ and the adjacent dimers are connected by the inter-cell matrix $T$. The nonvanishing elements of the intra-unit cell matrix $H$ are given by $H_{11} = H_{44} = \beta_1$, $H_{22} = H_{33} = \beta_2$, $H_{12} = H_{21} = H_{34} = H_{43} = \tilde{\kappa}$, and $H_{23} = H_{32} = \kappa$, while the inter-cell coupling matrix $T$ has only one non-zero entry $T_{14} = \kappa$. Without any loss of generality, we assume that $\beta_1 > \beta_2$ and take all the coupling coefficients to be real.

The topological bowtie lattice provides control to design different Bloch eigenstates formed through hybridization of the supermodes associated with the dimers/waveguides. As shown in Figure 1(b), for a nonvanishing detuning $\Delta\beta$ of the propagation constants between the large and small waveguides in each of the dimer, the supermodes are highly localized in the large (base) or small (vertex) waveguides. Two supermodes each are close to resonance, experiencing effective coupling strengths alternating between strong and weak. This effect can be viewed as two SSH Hamiltonians (SSH $_I$ and SSH $_{II}$) occupying the same space yet having independent topological numbers as shown in the lower panel of Figure 1(b). As each SSH model creates two eigenvalues $\lambda_\pm$, the designed bowtie lattice is expected to demonstrate four dispersion bands, which is confirmed by direct modelling (**Figure 1(c)**). The coupling strengths and propogation constants are engineered to demonstrate different topological phases and thus realize the related edge states. In the top panel of Figure 1c, the design parameters were chosen to be $\tilde{\kappa} = 0.127 \ \mu m^{-1}$, $\kappa = 0.5\tilde{\kappa}$ and $\beta_1 = -\beta_2 = \tilde{\kappa}$. As expected from the discussion above, the system resembles two separate SSH Hamiltonians giving rise to two upper (SSH $_I$) and two lower (SSH $_{II}$) bands. The two isolated eigenvalues in the spectrum (one in the upper and the other in the lower band gaps) correspond to states localized at the left and right edge. This is distinguished from the conventional SSH model for which the two edge states would lie in the same gap. The middle and lower panels of Figure 1(c) highlight a crucial novel feature of our model—the existence of a third, central gap that separates the two effective SSH models. The design parameters are the same as used in the top panel, but for $\kappa = \sqrt{2}\tilde{\kappa}$ and $2\tilde{\kappa}$, respectively. The central gap closes at $\kappa = \sqrt{2}\tilde{\kappa}$ while in the lower panel the gap is again opened but with a band inversion. This band inversion gives rise to an additional pair of isolated eigenvalues, which are accompanied by the emergence of two new edge states. These edge states are associated with the spectral symmetry of the bowtie chain, which induces an additional topological number. In an infinite long chain, the two low-energy solutions near $k = 0$ give rise to two slowly varying fields that can be grouped into a spinor $\varphi$. Its evolution takes the form of a Jackiw-Rebbi model $id\varphi/dz = H_{eff}\varphi$ with an effective Hamiltonian $H_{eff} = m\sigma_z + v_F\sigma_y\hat{p}_x$,[31] again in complete analogy with the SSH model.[32] All three effective models are therefore associated with a chiral symmetry $\sigma_x H_{eff} \sigma_x = -H_{eff}$ guaranteeing non-trivial topology in each gap.

The detailed edge features of the complete system can be understood by inspecting the Zak phase[33] and Witten index[34] associated with each bulk band and each bandgap, respectively (see [35-41] for more details). The Witten index is related to the reflection phase at a spectral symmetry point and can be calculated form the associated Zak phase of the bulk bands, which determine the reflection phases at the band edges. A detailed calculation for our setup [28] results in the relations $W_1 = -(Z_1 - Z_4)$ for the upper band gap, $W_2 = (Z_1 - Z_4) = (Z_2 - Z_3)$ for the lower band gap, and $W = -(Z_2 + Z_3)$ for the central band gap, where $Z_i$ are the Zak indices ordered from the top to the bottom band, as listed in **Table 1.** In configuration II the Zak phases of bands 2 and 3 are ill defined, so that there we adopt the Zak phases from just before the transition point. With the designed termination of the unit cell, we expect to find edge states in each gap at the right edge (lower panel of Figure 1(a)) when the corresponding Witten index takes the value $-1$, corresponding to fulfillment of the effective hard-wall boundary conditions. For the experiment, we exploit that the Witten index $W$ determining the existence of a topological edge state in the central bandgap can be controlled by solely tuning the intra-dimer coupling.

A straightforward way to demonstrate the topological features is to excite the system at its edge. In our photonic lattice design, the initial state is set up as the mode of the outmost waveguide (the right edge of the 1D waveguide lattice) where its propagation constant is fixed. By coupling light to one edge of the topological photonic lattice, all existing edge states at that edge will be excited because of the large overlap with the input state. We aim to identify the different topological phases through the discrete diffraction, localization and interference signatures of light transport. This is greatly facilitated when the light transport dynamics becomes directly visualized over the whole propagation distance, revealing the evolution of the transverse light distribution for a propagation distance *z*. To measure the light transport in the far field, we intentionally introduced periodic hole patterns on top of the waveguide lattice, satisfying a phase matching condition to coincide with the effective wavelength of the guided mode propagating inside of the waveguides and to efficiently couple a small proportion of the light into the upward direction, as illustrated in **Figure 2(a)**. For the experiments, we fabricated three different samples of the photonic lattice with controlled physical parameters corresponding to different configurations I, II and III as defined in **Table 2.** On an SOI platform, the samples were patterned using electron beam lithography, followed by reactive ion etching to form the bowtie waveguides lattice with the hole patterns. The diameter of each hole was chosen to be 150 nm, which provided a good balance between the upward coupling efficiency and the insertion loss. The $SiO_2$ cladding layer was subsequently deposited using plasma enhanced chemical vapor deposition (PECVD), which ensures symmetric confinement of the light field inside the waveguides and increases the efficiency of upward coupling. Each sample consisted of 18 guiding channels. The scanning electron microscope pictures before deposition of the $SiO_2$ cladding are shown in **Figure 2(b)**.

The effective realization of the edge states in different topological phases was both experimentally and numerically validated through the imaging of the light transport at the sample plane in three different configurations. A tunable continuous-wave fiber laser (adjusted to operate at the free space wavelength of 1555 nm) was directly connected to a polarization-maintaining tapered fiber that efficiently delivered a TM polarized laser beam into the right-most edge waveguide of the on-chip bowtie waveguide lattice. Figure 3(a) corresponds to configuration I ($\kappa = 0.5\tilde{\kappa} = 0.064\mu m^{-1}$), where our theoretical model predicts a single edge state confined in the bottom waveguide, originating from the Witten index $W_1 = -1$ for the upper finite-energy bandgap; this ensures a single topological edge state in that gap. Meanwhile, the Witten indices in the central and lower bandgaps are designed to be $W = W_2 = 1$, leading to no topological edge state at the bottom waveguide in these gaps. The optical intensity remains well confined to the launching channel (i.e. the bottom waveguide), while close inspection shows the absence of any appreciable intensity fluctuations. For configuration II ($\kappa = \sqrt{2}\tilde{\kappa} = 0.180\mu m^{-1}$) (Figure 3(b)), light localization at the edge persists as the Witten index for the finite energy gaps remains the same as that in configuration I. However, a clear signature of discrete diffraction across all the waveguides in the transverse direction is also observed, conforming with the general expectations for the closure of the central bandgap, which results in a nearly linear band-dispersion [30] that facilitates the observed secondary emission. From our modelling, the overlap between the input states and the bulk states becomes maximal at this point in Figure 3(d). Increasing the coupling $\kappa$ to the value in configuration III ($\kappa = 2\tilde{\kappa} = 0.255\mu m^{-1}$) leads to a reopening and inversion of the central bandgap. The Witten index of the central bandgap switches from $+1$ to $-1$, and fulfills the boundary condition [28] to form an edge state also inside this bandgap. Since the Witten indices $W_1$ and $W_2$ for the finite energy bandgaps remain unchanged, the system now support two edge states located in different bandgaps, and the distinct propagation constants of these two states leads to interference beating along the launch waveguide (Figure 3(c)). Since the reopened central bandgap is not as wide as that in configuration I, the corresponding edge state resides close to the band edge, such that there also exists pronounced diffraction into the bulk.

In the experiment some small intensity fluctuations are also observed in configurations I and II (but much smaller if compared with case III). This is due to the small overlaps between the input state and the extended bulk states (Figure. 3(d)) as well as some disorder (see Supporting Information for details). Some of these fluctuations result from the resolution of approximately 1 μm in our far-field imaging system, which thereby also captures light from the adjacent waveguides. To address this issue, we applied the time-resolved spatial-heterodyne imaging technique,[42] which provides the spatial distribution of the light versus time. Thus, we could characterize the ultrafast transport dynamics in the observed edge states. Here, a femtosecond pulse with a pulse width of ~840 fs centered at 1550 nm was delivered via the edge waveguide to propagate in the waveguide lattices. The apparatus was based on a modified Mach-Zehnder interferometer with a variable delay line, by which the ultra-fast time-resolved test can be performed by capturing interferograms corresponding to different time delays (see Supporting Information for more details). Consistent with the continuous wave measurements, the temporal evolution of the wave packet in the bowtie lattice further confirms the topological transition among the three designed configurations (Fig. 4). For configuration I, the launched wave packet couples mainly into a single edge state. Since the edge state is localized within the wide central bandgap, the pulse propagation is robust against variations of the neighboring couplings and thus remains confined in the right-most waveguide (see **Movie S1** in Supporting Information). The intensity variations in the neighboring waveguides can be understood to result from the mismatch between the excitation in a single waveguide and the actual modal profile of the edge state, which extends over a few waveguides. These ultrafast temporal measurements provide access to quantitative characteristics of the edge state.[43]

All the edge states are associated with distinct dynamical properties encoded in the effective group and phase indices, which provide additional quantitative assessments of each state. In configuration I, the group index $n_g = 3.30 \pm 0.012$ (calculation details can be found in supporting information) can be retrieved through pulse positions traveled at different time delays, corresponding to an effective index of $n_{eff} = 1.67 \pm 0.012$ that agrees well with the simulation $n_{eff} = 1.72$. For configuration II, it is clearly demonstrated that the dynamical transport of the single edge state is accompanied by a secondary emission, revealing the closure of the central bandgap. Their interference, while weak, slightly distorts the field distribution and the propagation of the wave packet in the launching channel (see **Movie S2** in Supporting Information). The measured group index is consistently lower than configuration I, $n_g = 3.23 \pm 0.011$ with $n_{eff} = 1.66 \pm 0.011$. For configuration III, the dynamical evolution of the wave packet is revealed by the interference beating [44-46] due to the co-propagation of two edge states with distinct propagation constants (see also **Movie S3** in Supporting Information). The measured group index $n_g = 3.18 \pm 0.015$ in this case is the averaged group index of the two edge states. Their respective effective indices are $n_{eff,1} = 1.70 \pm 0.015$ and $n_{eff,2} = 1.65 \pm 0.015$. In contrast, a uniformly arranged trivial waveguide array shows a diffraction pattern corresponding to free spreading and reflection of the wave packet across the whole array. This is distinct from the topological edge modes observed in the previous 3 configurations. (See **Movie S4** in the Supporting Information.)

While the samples are designed for excitation at the outmost waveguide, observation of the diffraction from the edge to the bulk state also provides convincing evidences to judge if a band gap is closed or open. Meanwhile, the appearance of the beating patterns directly reveals the engagement of a second edge mode, which arises in the newly opened second topological bandgap. This configuration with multiple topological bandgaps is in contrast with recent work where two topological edge states emerge through band folding in the same bandgap.[47] In our case, both the continuous wave and temporally resolved experimental results confirm that the multiple topological numbers of out photonic lattice offer more flexible control over the topological states.

In summary, by considering the non-trivial square root of a two legged ladder system, we designed and experimentally demonstrated a versatile photonic lattice with multi-band topology. Compared with the conventional Su-Schrieffer Heeger and Rice-Mele models, the lattice offers additional spectral symmetries that enrich the topological features and enable to induce independently tuned edge states. We experimentally investigated the ultrafast beam transport dynamics to validate the supported topological characteristics. Through systematically manipulating the couplings in the lattice, the topological nature of multiple dispersion bands can be effectively engineered with a desired Witten index in different energy bandgaps, enabling the versatile realization of topologically-induced edge state dynamics.

**Acknowledgements**
Z.Z., M.P., P.M., and L.F. acknowledge support from the Army Research Office Young Investigator Research Program (Grant No. W911NF-16-1-0403) and King Abdullah University of Science & Technology (Grant No. OSR-2016-CRG5-2950-04). This research was partially supported by NSF through the University of Pennsylvania Materials Research Science and Engineering Center (MRSEC) (DMR-1720530). This work was carried out in part at the Singh Center for Nanotechnology, part of the National Nanotechnology Coordinated Infrastructure Program, which is supported by the NSF grant NNCI-1542153. M.H.T. and R.E.-G. acknowledge support from the Army Research Office (W911NF-17-1-0481) and the National Science Foundation (Grant No. ECCS-1545804). J.A. and H.S. acknowledge support from EPSRC (Grant No. EP/L01548X/1 and Grant No. EP/N031776/1).

**References**

[1] R. B. Laughlin, *Phys. Rev. B* **1981**, 23, 5632.

[2] D. J. Thouless, M. Kohmoto, M. P. Nightingale, M. Den Nijs, *Phys. Rev. Lett.* **1982**, 49, 405.

[3] F. D. M. Haldane, *Phys. Rev. Lett.* **1988**, 61, 2015.

[4] D. N. Christodoulides, R. I. Joseph, *Optics Lett.* **1988**, 13, 794-796.

[5] M.C. Rechtsman, J.M. Zeuner, Y. Plotnik, Y. Lumer, D. Podolsky, F. Dreisow, S. Nolte,

M. Segev, A. Szameit, *Nature* **2013**, 496, 196-200.

[6] M. Hafezi, E. A. Demler, M. D. Lukin, J. M. Taylor, *Nat. Phys.* **2011**, 7, 907-912.

[7] M. Hafezi, S. Mittal, J. Fan, A. Migdall, J. M. Taylor, *Nat. Photon.* **2013**, 7, 1001-1005.

[8] G. Q. Liang, Y. D. Chong, Phys. *Rev. Lett.* **2013**, 110, 203904.

[9] M.C. Rechtsman, Y. Plotnik, J.M. Zeuner, D. Song, Z. Chen, A. Szameit, M. Segev, Phys. *Rev. Lett.* **2013**, 111, 103901.

[10] S. Weimann, M. Kremer, Y. Plotnik, Y. Lumer, S. Nolte, K.G. Makris, M. Segev, M.C. Rechtsman, A. Szameit, *Nat. Mater.* **2017**, 16, 433-438.

[11] Z. Wang, Y. D. Chong, J. D. Joannopoulos, M. Soljačić, *Nature* **2009**, 461, 772-775.

[12] A. B. Khanikaev, S. H. Mousavi, W. Tse, M. Kargarian, A. H. MacDonald, G. Shvets, *Nat. Mater.* **2013**, 12, 233-239.

[13] L. H. Wu, X. Hu, *Phys. Rev. Lett.* **2015**, 114, 223901.

[14] C. He, X. C. Sun, X. P. Liu, M. H. Lu, Y. Chen, L. Feng, Y. F. Chen, *PNAS* **2016**, 113, 4924-4928.

[15] B. Zhen, C. W. Hsu, L. Lu, A. D. Stone, M. Soljačić, *Phys. Rev. Lett.* **2014**, 113, 257401.

[16] S. A. Skirlo, L. Lu, Y. Igarashi, Q. Yan, J. Joannopoulos, and M. Soljačić, *Phys. Rev. Lett.* **2015**, 115, 253901.

[17] A. Kodigala, T. Lepetit, Q. Gu, B. Bahari, Y. Fainman, B. Kanté, *Nature* **2017**, 541, 196-199.

[18] M. Verbin, O. Zilberberg, Y. E. Kraus, Y. Lahini, Y. Silberberg, *Phys. Rev. Lett.* **2013**, 110, 076403.

[19] M. A. Bandres, M. C. Rechtsman, M. Segev, *Phys. Rev. X* **2016**, 6, 011016.

[20] H. Schomerus, *Opt. Lett.* **2013**, 38, 1912-1914.

[21] C. Poli, M. Bellec, U. Kuhl, F. Mortessagne, H. Schomerus, *Nat. Commun.* **2015**, 6, 6710.

[22] H. Zhao, P. Miao, M. H. Teimourpour, S. Malzard, R. El-Ganainy, H. Schomerus, L. Feng, *Nat. Commun.* **2018**, 9, 981.

[23] M. Parto, S. Wittek, H. Hodaei, G. Harari, M. A. Bandres, J. Ren, M. C. Rechtsman, M. Segev, D. N. Christodoulides, M. Khajavikhan, *Phys. Rev. Lett.* **2018**, 120, 113901.

[24] P. St-Jean, V. Goblot, E. Galopin, A. Lemaître, T. Ozawa, L. Le Gratiet, I. Sagnes, J. Bloch, A. Amo, *Nat. Photon.* **2017**, 11, 651.

[25] G. Harari, M. A. Bandres, Y. Lumer, M. C. Rechtsman, Y. D. Chong, M. Khajavikhan, D. N. Christodoulides, M. Segev, *Science* **2018**, 359, 6381.

[26] M. A. Bandres, S. Wittek, G. Harari, M. Parto, J. Ren, M. Segev, D. N. Christodoulides, M. Khajavikhan, *Science* **2018**, 359, 6381.

[27] W. Su, J. R. Schrieffer, A. J. Heeger, *Phys. Rev. Lett.* **1979**, 42, 1698.

[28] J. Arkinstall, M. H. Teimourpour, L. Feng, R. El-Ganainy, H. Schomerus, *Phys. Rev. B* **2017**, 95, 165109.

[29] M. J. Rice, E. J. Mele, *Phys. Rev. Lett.* **1982**, 49, 1455.

[30] D. N. Christodoulides, F. Lederer, Y. Silberberg, *Nature* **2003**, 424, 817.

[31] R. Jackiw, C. Rebbi, *Phys. Rev. D* **1976**, 13, 3398.

[32] S. Ryu, Y. Hatsugai, *Phys. Rev. Lett.* **2002**, 89, 077002.

[33] J. Zak, *Phys. Rev. Lett.* **1989**, 62, 2747.

[34] E. Witten, *Nucl. Phys. B* **1982**, 202, 253-316.

[35] M. Xiao, Z.-Q. Zhang, C.-T. Chan, *Phys. Rev. X* **2014**, 4, 021017.

[36] X. Shi, C. Xue, H. Jiang, H. Chen, *Opt. Express* **2016**, 24, 18580.

[37] D. Meidan, T. Micklitz, P. W. Brouwer, *Phys. Rev. B* **2011**, 84, 195410.

[38] I. C. Fulga, F. Hassler, A. R. Akhmerov, *Phys. Rev. B* **2012**, 85, 165409.

[39] A. J. Niemi, G. W. Semenoff, *Phys. Rep.* **1986**, 135, 99.

[40] D. Bolle, F. Gesztesy, H. Grosse, W. Schweiger, B. Simon, *J. Math. Phys.* **1987**, 28,

1512.

[41] N. V. Borisov, W. Muller, R. Schrader, *Commum. Math. Phys.* **1988**, 114, 475.

[42] R. Rokitski, K. A. Tetz, Y. Fainman, *Phys. Rev. Lett.* **2005**, 95, 177401.

[43] M. Abashin, K. Ikeda, R. Saperstein, Y. Fainman, *Opt. Lett.* **2009**, 34, 1327-1329.

[44] R. Rokitski, P. C. Sun, Y. Fainman, *Opt. Lett.* **2001**, 26, 1125-1127.

[45] L. Feng, M. Ayache, J. Huang, Y.L. Xu, M.H. Lu, Y.F. Chen, Y. Fainman, A. Scherer, *Science* **2011**, 333, 729-733.

[46] A. Regensburger, M.A. Miri, C. Bersch, J. Näger, Onishchukov, G. D.N. Christodoulides, U. Peschel, *Phys. Rev. Lett.* **2013**, 110, 223902.

[47] S. Yves, R. Fleury, T. Berthelot, M. Fink, F. Lemoult, G. Lerosey, *Nat. Commun.* **2017**, 8, 16023.

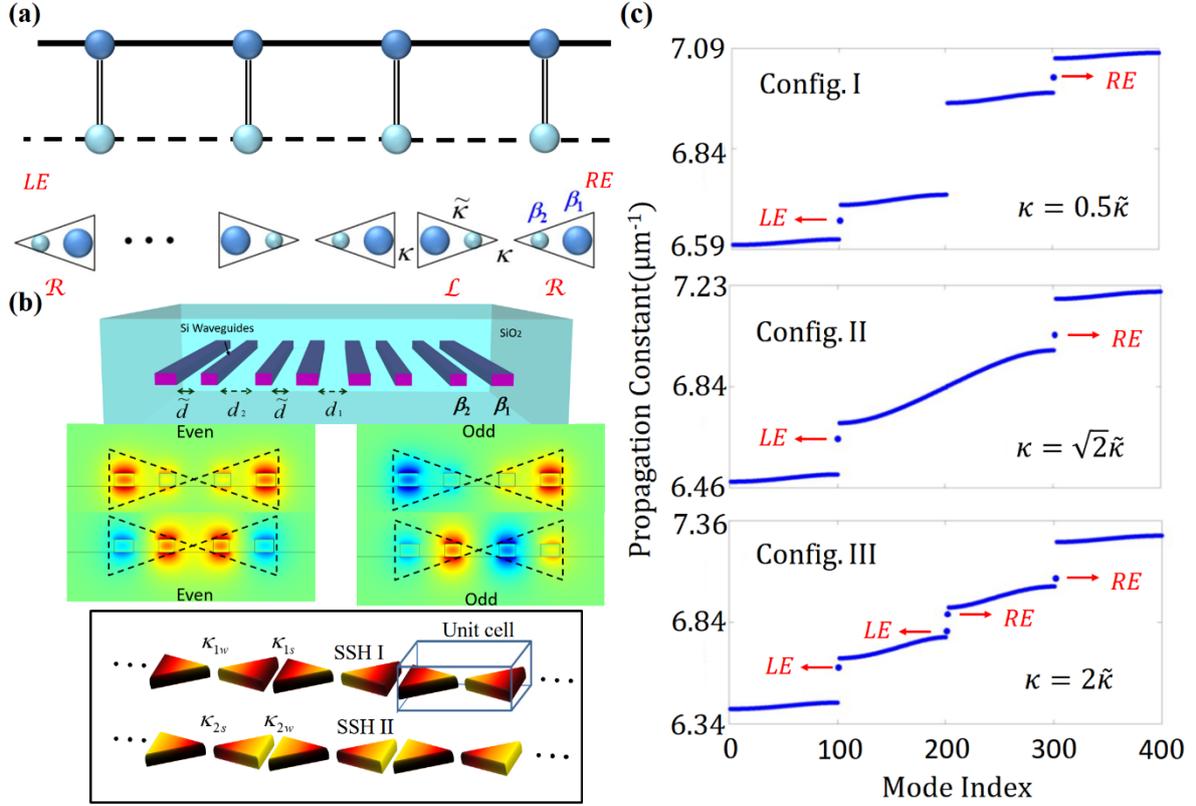

**Figure 1. Bowtie topological lattice.** (a) Two-legged ladder model having identical sites (top panel). Single and double lines represent couplings of different strength; the dashed lines signify couplings of opposite sign from the solid lines. Taking the square root and a $Z_2$ gauge transformation of this model results in the bowtie chain shown in the lower panel, with alternating couplings $\kappa$, $\tilde{\kappa}$ and staggered sequence of onsite energies $\beta_1, \beta_2, \beta_2, \beta_1, \ldots$. As indicated, this can be interpreted as a sequence of oppositely orientated dimers, labelled by L and R. The left edge of the chain is marker as LE and the right edge is marked as RE. (b) Implementation of the bowtie lattice using silicon waveguides embedded in silica cladding (only two unit cells are shown), where fundamental $TM_{00}$ mode hybridizations are formed for an isolated unit cell. The waveguide array consists of two different types of waveguides, having the same height $h = 230$nm but different widths: $w_1 = 300$nm and $w_2 = 350$nm, corresponding to propagation constants $\beta_1 = 6.713 \mu m^{-1}$ and $\beta_2 = 6.968 \mu m^{-1}$, respectively. These parameters translate into an onsite detuning of $\Delta\beta = 0.255 \mu m^{-1}$. In our design, different

waveguides are arranged in pairs with having $\beta_1$ and $\beta_2$, with a fixed separation $\tilde{d} = 475$nm that corresponds to an intra-dimer coupling $\tilde{\kappa} = 0.127 \mu m^{-1}$. The inter-dimer distances between the two types of waveguide are denoted by $d_{1,2}$, and are tuned to yield an identical coupling of $\kappa$. Lower panel shows the formation of two independent SSH Hamiltonians as a result of the eigenstate hybridization between the local modes of the $R$ and $L$ dimers. In SSH$_I$, the states are more localized at the bases giving rise to alternating strong and weak coupling at the bases/vertices respectively, and the converse for SSH$_{II}$. $\kappa_{s,w}$ indicate strong/weak coupling between the supermodes of each dimer. (c) Band structures of bowtie arrays of the form $(RL)_{100}R$ with the designed parameters. For $\kappa = 0.5\tilde{\kappa}$ (top pannel), the upper and lower bandgaps each support a single defect edge state, one on each edge. As the coupling reaches $\kappa = \sqrt{2}\tilde{\kappa}$ (middle panel), the two inner bands merge, closing the central bandgap. Further increasing the coupling to $\kappa = 2\tilde{\kappa}$ (bottom panel) the central gap is open again, which results in the emergence of two new edge states associated with an effective SSH model for the central gap. Edge states are marked with LE or RE to indicate their residence edge.

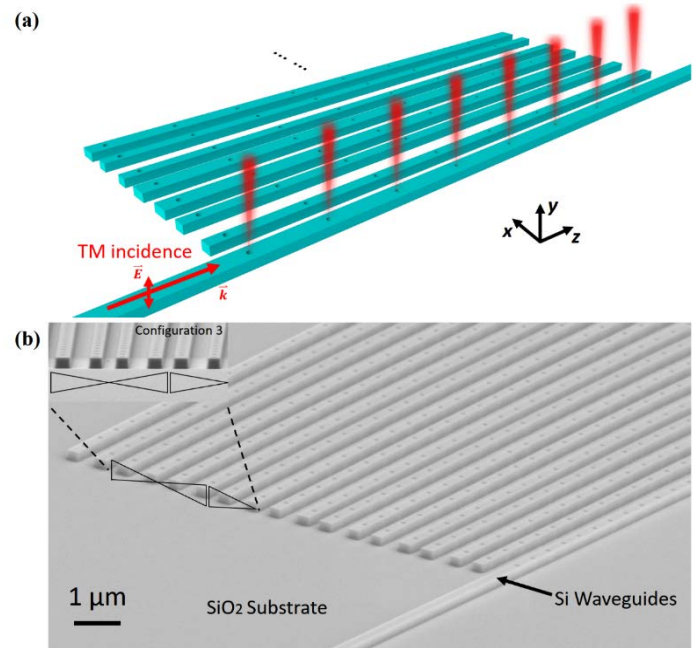

**Figure 2. Experimental implementation of bowtie topological photonic lattices.** (a) Configuration of the bowtie waveguide array with periodic hole patterns on an SOI platform, designed to measure the beam propagation dynamics across the length of the device. The on-top holes with a diameter of 150 nm are designed to couple light out of the waveguides, which extract light to free space to reveal the propagation of the light inside the structure in far field. (b) Scanning electron microscope pictures of the device (configuration III) before deposition of the $SiO_2$ cladding. The fabricated device consists of 18 waveguides. The periodicity of the holes is matched to the effective wavelength of the quasi-TM mode inside the waveguides (934 nm for the base waveguides and 901 nm for the vertex waveguides), which results in vertical light extraction. Inset zoom in the picture shows the cross section of the bowtie waveguides structure in configuration III.

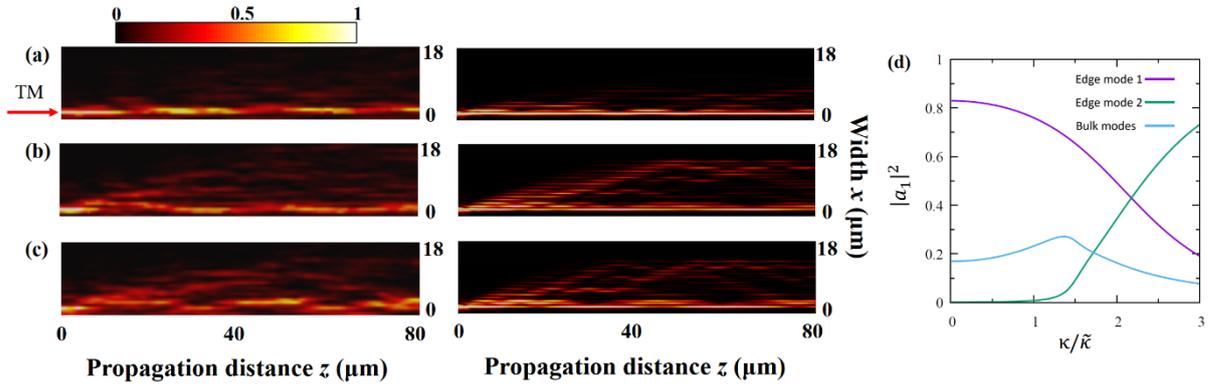

**Figure 3. Experimental and EM simulation beam transport in the bowtie topological waveguide lattices and modal overlap of initial state (delta excitation at the bottom waveguide) with the different supermodes of the system (edge-defect and bulk states).** (a)-(c) Light field intensity images under TM polarized continuous wave incidence at a wavelength of 1555 nm for configurations I, II and III (top to bottom panels), clearly demonstrating the transition between different topological phases. Due to the insertion loss arising from the hole array, the total propagating power across the bowtie waveguides lattice slightly decreases as a function of the propagation distance. Normalizing the recorded images with respect to the total power across every propagation cross section, so that the total power at any distance $z$ remains a constant, enables a fair comparison between the experimental results (left panels) and simulations (right panels). (d) Predicted modal overlap of the initial input state with the edge states and the bulk states for the 18 waveguides experimental system, as a function of the hopping amplitude. $|a_1|^2$ describes the modal overlaps between the input and excited states. The modal overlap between the input mode and the bulk mode, while weak compared to the overlap with edge states, leads to discrete diffraction into bulk and small intensity fluctuations observed in (a)-(c). Note that the same set of samples are used as for Figure 4. Images are zoomed in to show the details.

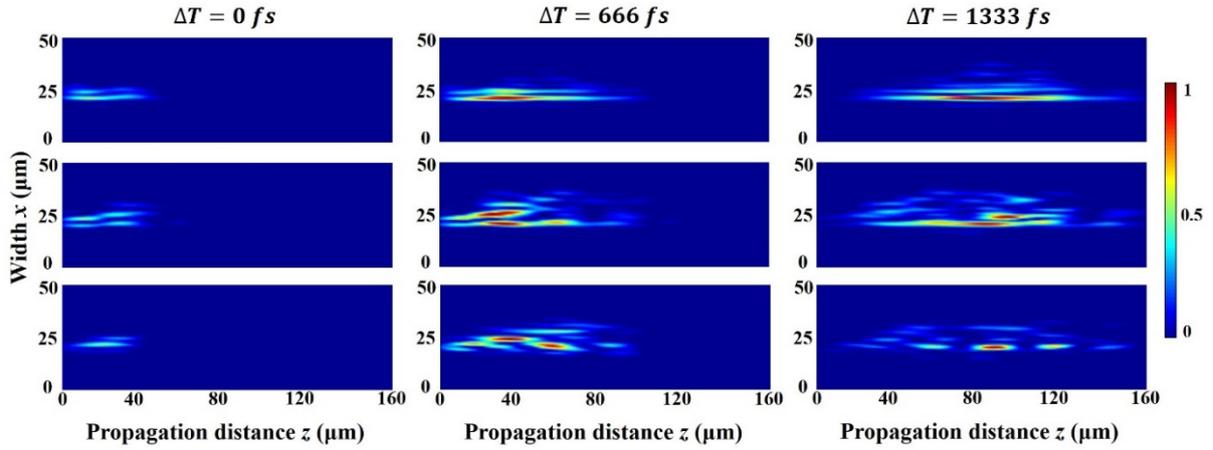

**Figure 4. Measured ultrafast transport dynamics in the bowtie topological waveguide lattices.** Temporal evolution of spatial intensity of the wave packet is captured with a time delay of ~66.6 fs for configurations I, II and III (top, middle, and bottom panels, respectively). Images are normalized with the same input power, assuming a lossless propagation in the z direction. Field intensity spatial maps in left, middle, and right colums correspond to different time delays at $\Delta T = 0, 666,$ and 1333 fs, respectively, showing the wave packet entering the lattice, the formation of the edge states at the beginning of the lattice, and the transport of the edge states in the lattice. More detailed information can be found in Movies S1, S2, and S3 in Supporting Information.

**Table 1**. Zak phase and Witten index of 3 configurations in different topological phases.

| Band/Gap | Upper Band 1 $Z_1$ | Upper Gap $W_1$ | Upper Band 2 $Z_2$ | Central Gap $W$ | Lower Band 1 $Z_3$ | Lower Gap $W_2$ | Lower Band 2 $Z_4$ |
|---|---|---|---|---|---|---|---|
| Config.I | $Z_1 = 0$ | $W_1 = -1$ | $Z_2 = 0$ | W=1 | $Z_3 = -1$ | $W_2 = 1$ | $Z_4 = -1$ |
| Config.II | $Z_1 = 0$ | $W_1 = -1$ | $Z_2 = 0$ | N.A. | $Z_3 = -1$ | $W_2 = 1$ | $Z_4 = -1$ |
| Config.III | $Z_1 = 0$ | $W_1 = -1$ | $Z_2 = 1$ | W=-1 | $Z_3 = 0$ | $W_2 = 1$ | $Z_4 = -1$ |

Table 2. Design parameters of 3 configurations in different topological phases.

| Edge to edge separation | Configuration I $\kappa = 0.5\tilde{\kappa} = 0.064 \mu m^{-1}$ | Configuration II $\kappa = \sqrt{2}\tilde{\kappa} = 0.180 \mu m^{-1}$ | Configuration III $\kappa = 2.0\tilde{\kappa} = 0.255 \mu m^{-1}$ |
|---|---|---|---|
| $d_1$ | 700nm | 430nm | 345nm |
| $d_2$ | 610nm | 365nm | 290nm |